\documentclass{cpc-hepnp}
\usepackage{multicol}
\usepackage{graphicx}
\usepackage{booktabs}
\usepackage{amssymb,bm,mathrsfs,bbm,amscd}
\usepackage[tbtags]{amsmath}
\usepackage{lastpage}

\begin{document}


\fancyhead[c]{\small  To be published in 'Chinese Physics C'} \fancyfoot[C]{\small 000000-\thepage}


\title{Production of the exotic neutron-deficient isotopes near $N, Z$ = 50 in multinucleon transfer reactions\thanks{Project supported by
the National Natural Science Foundation of China (Grant Nos.~11635003, 11025524 and 11161130520),
the National Basic Research Program of China (Grant No.~2010CB832903),
and the European Commissions 7th Framework Programme (FP7-PEOPLE-2010-IRSES) (Grant No.~269131).}}

\author{%
      Xin-Xin Xu$^{1,2}$, Gen Zhang$^{1,2}$, Jing-Jing Li$^{1,2}$,
Bing Li$^{1,2}$, Cheikh A. T. Sokhna$^{1,2}$, Xin-Rui Zhang$^{1,2}$, \\Xiu-Xiu Yang$^{1,2}$, Shi-Hui Cheng$^{1,2}$, Yu-Hai Zhang Zhi-Shuai Ge$^{1,2}$, Cheng Li$^{1,2}$£¬Zhong Liu$^{3}$
\\and Feng-Shou Zhang$^{1,2,4}$\email{Corresponding author: fszhang@bnu.edu.cn}
}
\maketitle

\address{%
$^{1}$Key Laboratory of Beam Technology of Ministry of Education,
\\College of Nuclear Science and Technology,
\\ Beijing Normal University, Beijing 100875, China\\
$^{2}$Beijing Radiation Center, Beijing 100875, China\\
$^{3}$Institute of Modern Physics, Chinese Academy of Sciences, Lanzhou 730000, China\\
$^{4}$Center of Theoretical Nuclear Physics,
\\National Laboratory of Heavy Ion Accelerator of Lanzhou, Lanzhou 730000, China
}

\begin{abstract}
The multinucleon transfer reaction in the collisions of $^{40}$Ca+$^{124}$Sn at $E_{\textrm{c.m.}}=128.5$ MeV is investigated by using the improved quantum molecular dynamics model. The measured
angular distributions and isotopic distributions of the products are reproduced reasonably well by the calculations. The multinucleon transfer reactions of $^{40}$Ca+$^{112}$Sn, $^{58}$Ni+$^{112}$Sn, $^{106}$Cd+$^{112}$Sn, and $^{48}$Ca+$^{112}$Sn are also studied. It shows that the combinations of neutron-deficient projectile and target are advantageous to produce the exotic neutron-deficient nuclei near $N, Z$ = 50. The charged particles emission plays an important role at small impact parameters in the deexcitation processes of the system.
The production cross sections of the exotic neutron-deficient nuclei in multinucleon transfer reactions are much larger
than those measured in the fragmentation and fusion-evaporation reactions. Several new neutron-deficient nuclei can be produced in $^{106}$Cd+$^{112}$Sn reaction. The corresponding production cross sections for the new neutron-deficient nuclei, $^{101,102}$Sb, $^{103}$Te, and $^{106,107}$I, are 2.0 nb, 4.1 nb, 6.5 nb, 0.4 $\mu$b and 1.0 $\mu$b, respectively.
\end{abstract}



\footnotetext[0]{\hspace*{-3mm}\raisebox{0.3ex}{$\scriptstyle\copyright$}2013
Chinese Physical Society and the Institute of High Energy Physics
of the Chinese Academy of Sciences and the Institute
of Modern Physics of the Chinese Academy of Sciences and IOP Publishing Ltd}%


\section{Introduction}

The production of the neutron-deficient nuclei near $N, Z$ = 50 has attracted extensive attention in recent years \cite{Sn1,Sn2,Sn3,Sn4,Sn5,Sn6}. On the one hand, the structure and decay properties of these nuclei play special roles in nuclear physics. For the nuclei with $N\approx Z$ around the doubly magic nucleus $^{100}$Sn, the valence neutrons and protons occupy the same orbits. Therefore, the neutron-proton pairing correlations would be enhanced greatly. The level structure of the nuclei below $^{100}$Sn is described quite well in the basic shell model space with a rigid core at $N=Z=50$ and valence holes. The core-excited states have been observed in semimagic nuclei including $^{95}$Rh, $^{96}$Pd, $^{97}$Ag, $^{98}$Cd, etc \cite{core1,core2,core3}. The decay properties of the neutron-deficient nuclei around $^{100}$Sn have revealed interesting phenomena, such as the Gamow-Teller transition, proton decay, two-proton decay, $\alpha$ decay, etc \cite{Gam1,alpha1,alpha2,alpha3}. In addition, the clusters $(Z\geq3)$ emission from the excited state of nuclei has been observed in this region. On the other hand, the rp-process (a sequence of proton captures followed by $\beta^+$ decays) runs along the $N=Z$ line, which is believed to be responsible for the production of the stable isotopes on the neutron-deficient side of the valley of $\beta$ stability. To produce these nuclei is also extremely important for understanding the rp-process in nuclear astrophysics \cite{rp1,rp2,rp3}.

The fragmentation reactions of $^{238}$U at intermediate and relativistic energies could produce the new neutron-rich nuclei with a wide range \cite{frag1,frag2,frag3,frag4}. However, the new neutron-deficient nuclei near $^{100}$Sn could be obtained through the fragmentation of neutron-deficient projectile, such as $^{106}$Cd, $^{112}$Sn, and $^{124}$Xe. Many such experiments were performed at RIKEN \cite{Sn6}, GSI \cite{GSI}, MSU \cite{MSU}, and GANIL \cite{GANIL}. For example, four new nuclei, $^{96}$In, $^{94}$Cd, $^{92}$Ag, and $^{90}$Pd, were observed with cross sections about $10^{-12}-10^{-11}$mb by using a 345 MeV/A $^{124}$Xe beam impinging on a Be target at RIKEN \cite{Sn6}. The $^{100}$Sn was also identified in this experiment with production cross section about $10^{-9}$ mb. Another method to produce the nuclei near $^{100}$Sn is fusion-evaporation reaction of stable neutron-deficient partners, such as $^{63}$Cu($^{40}$Ca,3n)$^{100}$In \cite{fus1}, $^{58}$Ni($^{50}$Cr,$\alpha$p3n)$^{100}$In \cite{fus2}, $^{58}$Ni($^{58}$Ni,$^{12}$C3n)$^{101}$Sn \cite{fus2}, and so on. The cluster and light charged particles emissions play an important role in the deexcitation processes of the neutron-deficient compound nucleus. In Ref. \cite{fus3}, $^{100}$Sn was observed in the fusion-evaporation reaction of 255 MeV $^{50}$Cr with $^{58}$Ni at GANIL. The reported production cross section of $^{100}$Sn was $10^{-5}$ mb. Recently, the reaction $^{54}$Fe($^{58}$Ni,4n)$^{108}$Xe was performed at the ATLAS facility of Argonne National Laboratory using the Fragment Mass Analyzer (FMA). The self-conjugate $^{108}$Xe $\rightarrow$ $^{104}$Te $\rightarrow$ $^{100}$Sn $\alpha$-decay chain was observed for the first time \cite{fus4}.

The multinucleon transfer (MNT) reaction can be used to generate the exotic nuclei far away from the $\beta$ stability line, including not only the neutron-rich nuclei, but also the neutron-deficient nuclei. For example, the new neutron-deficient nuclei, $^{216}$U, $^{219}$Np, $^{223}$Am, $^{229}$Am, and $^{233}$Bk, were produced in the MNT reaction of $^{48}$Ca+$^{248}$Cm at GSI \cite{MNT1}. In Ref. \cite{MNT2}, another MNT reaction of $^{136}$Xe+$^{198}$Pt at 7.98 MeV/A showed that the production cross sections of the neutron-rich nuclei with $N=126$ are much larger than those measured in the fragmentation reaction of $^{208}$Pb+$^{9}$Be at 1 GeV/A \cite{pb}. Besides the experimental achievements, some models have been developed to describe the MNT processes during the recent years. The ImQMD model is a semiclassical microscopic dynamics model based on effective nucleon-nucleon interaction, which is successfully applied to heavy-ion collisions at intermediate and low energies \cite{low1,low2,low5}. Other models for describing the MNT reactions include the Complex WKB (CWKB) theory \cite{CWKB1,CWKB2}, the isospin-dependent quantum molecular dynamics model (IQMD) \cite{IQMD1,IQMD2,IQMD3,IQMD4,IQMD5}, the dinuclear system (DNS) model \cite{DNS1,DNS2,DNS3,DNS4,DNS5,DNS6,DNS7,DNS8} and so on.
For a comprehensive review, please read Ref. \cite{FOP}.

In this work, we attempt to produce the neutron-deficient nuclei near $N, Z$ = 50 with the MNT reactions by using the ImQMD model. The structure of this paper is as follows. In Sec. II, we briefly introduce the ImQMD model. The results and discussion are presented in Sec. III. Finally the conclusion is given in Sec. IV.



\section{Theoretical framework}

In the ImQMD model, the same as the original QMD model \cite{QMD3}, each nucleon is represented by a coherent state of a Gaussian wave packet
\begin{equation}  \label{1}
\phi _{i}(\mathbf{r})=\frac{1}{(2\pi \sigma _{r}^{2})^{3/4}}\exp [-\frac{(
\mathbf{r-r}_{i})^{2}}{4\sigma
_{r}^{2}}+\frac{i}{\hbar}\mathbf{r}\cdot \mathbf{p}_{i}],
\end{equation}
where $\mathbf{r}_{i}$ and $\mathbf{p}_{i}$ are the centers of $i$th wave
packet in the coordinate and momentum space, respectively. $\sigma
_{r}$ represents the spatial spread of the wave packet in the coordinate space.
The time evolution of $\mathbf{r}_{i}$ and $\mathbf{p}_{i}$ for each nucleon is governed by Hamiltonian equations of motion
\begin{equation}  \label{8}
\dot{\mathbf{r}}_{i}=\frac{\partial H}{\partial \mathbf{p}_{i}}, \dot{%
\mathbf{p}}_{i}=-\frac{\partial H}{\partial \mathbf{r}_{i}}.
\end{equation}
The Hamiltonian of the system includes the kinetic energy $T=\sum\limits_{i} \frac{\mathbf{p}_{i}^{2}}{2m}$ and effective
interaction potential energy
\begin{equation}  \label{9}
H=T+U_{\mathbf{Coul}}+U_{\mathbf{loc}},
\end{equation}
where, $U_{\mathbf{Coul}}$ is the Coulomb energy, which is written as a sum of the direct and
the exchange contribution
\begin{eqnarray}
 \nonumber U_{\mathbf{Coul}}=&\frac{1}{2}\int\int{\rho_{p}(\textbf{r})}
 \frac{e^{2}}{|\textbf{r}-\textbf{r}'|}{\rho_{p}(\textbf{r}')}d\textbf{r}d\textbf{r}'\\
 &-e^{2}\frac{3}{4}(\frac{3}{\pi})^{1/3}\int\rho_{p}^{4/3}d\textbf{r}.\label{16}
\end{eqnarray}
Here, $\rho_{p}$ is the density distribution of protons of the system. The nuclear interaction potential energy $U_{\textrm{loc}}$ is obtained from the integration of the Skyrme energy density functional $U=\int V_{\textrm{loc}}(\mathbf{r})d\mathbf{r}$ without the spin-orbit term, which reads
\begin{eqnarray}
\nonumber V_{\mathbf{loc}}=&&\frac{\alpha }{2}\frac{\rho ^{2}}{\rho _{0}}+\frac{\beta }{\gamma +1}%
\frac{\rho ^{\gamma +1}}{\rho _{0}^{\gamma }}+\frac{\textsl{g}_{sur}}{2\rho _{0}}%
(\nabla \rho )^{2}\\
&&\ +\frac{C_{s}}{2\rho _{0}}(\rho ^{2}-\kappa _{s}(\nabla \rho
)^{2})\delta ^{2} + g_{\tau}\frac{\rho ^{\eta +1}}{\rho_{0}^{\eta
}}. \label{12}
\end{eqnarray}
Here, $\rho=\rho_{n}+\rho_{p}$ is the nucleons density. $\delta=(\rho_{n}-\rho_{p})/(\rho_{n}+\rho_{p})$ is the isospin asymmetry. The first three terms in above expression are obtained from the Skyrme interaction directly. The fourth term is the symmetry potential energy including the bulk and the surface symmetry potential energies. The last term is a small correction term. The parameters named IQ2 (see Table I) adopted in this work have been tested for describing the fusion reactions \cite{QMD4}, MNT reactions \cite{low1,low2,low5,QMD6}, and fragmentation reactions \cite{MD1}. The phase space occupation constraint method proposed by Papa et al. in the constrained molecular dynamics (CoMD) model \cite{Papa} is adopted to describe the fermionic nature of the $N$-body system. It improves greatly the stability of an individual nucleus.

\begin{table*}
\tabcolsep=15pt \caption{The model parameters (IQ2) adopted in this work.}
{\begin{tabular}{@{}cccccccccc@{}}

\hline\hline
   $\alpha$ & $\beta$ & $\gamma$ & $\textsl{g}_{sur}$ & $\textsl{g}_{\tau}$ & $\eta$ & $C_{S}$ & $\kappa_{s}$ & $\rho_{0}$ \\
  $(\textrm{MeV})$ & $(\textrm{MeV})$ & $$ & $(\textrm{MeV}\cdot \textrm{fm}^{2})$ & $(\textrm{MeV})$ & $$ & $(\textrm{MeV})$ & $(\textrm{fm}^{2})$ & $(\textrm{fm}^{-3})$ \\
\hline
   $-356$ & $303$ & $7/6$ & $7.0$ & $12.5$ & $2/3$ & $32.0$ & $0.08$ & $0.165$\\
\hline\hline
\end{tabular}}

\end{table*}

In this work, we set $z$-axis as the beam direction and $x$-axis as the impact parameter direction. The initial distance of the center of mass between the projectile and target is 30 fm. The wave-packet width is set as $\sigma_r=1.2$ fm. The dynamic simulation is stopped at 1000 fm/$c$. And then the GEMINI code \cite{GEMI1,GEMI2} is used to deal with the subsequent de-excitation process. The evaporation of the light particles is treated by the Hauser-Feshbach theory \cite{hua} including $n$, $p$, $d$, $t$, $^3$He, $\alpha$, $^6$He, $^{6-8}$Li, and $^{7-10}$Be channels. The level density in GEMINI code, is obtained by the Fermi gas expression
\begin{equation}  \label{6}
\rho(U,J)=(2J+1)[\frac{\hbar^{2}}{2\mathscr{I}}]^{3/2}\frac{\sqrt{a}}{12}\frac{\exp(2\sqrt{aU})}{U^2},
\end{equation}
where the $\mathscr{I}$ is the moment-of-inertia of the residual nucleus or saddle-point configuration. The level density parameter was taken as a= A/8 MeV$^{-1}$ as usual.

\section{Results and discussion}

To test the ImQMD model for the description of the MNT reactions, we calculate the collisions of $^{40}$Ca+$^{124}$Sn at $E_{\textrm{c.m.}}=128.5$ MeV. The range of the impact parameters in the calculations is from 0 to $b_{\textrm{max}}$ fm. $b_{\textrm{max}}=R_\textrm{P}+R_\textrm{T}$, where $R_\textrm{P}$ and $R_\textrm{T}$ denote the radii of the projectile and target, respectively. The incident energy is slightly higher than the Coulomb barrier (120.1 MeV). For central collisions, most events are fusion reactions. The fusion and elastic scattering events are not taken into account in our analysis. Figure 1 shows that the angular distributions of the final projectile-like-fragments (PLFs) with different transfer channels in $^{40}$Ca+$^{124}$Sn at $E_{\textrm{c.m.}}=128.5$ MeV. The grazing angle in the laboratory frame is $75^\circ$. One sees that calculated maximum of the cross sections decreases more quickly with increasing neutron pickup channel than the experimental data \cite{CWKB1}. However, the positions of the maximum always keep consistent with the experimental data, which locate at the grazing angle with a small dependence on the channel.


\begin{center}
\includegraphics[width=12cm]{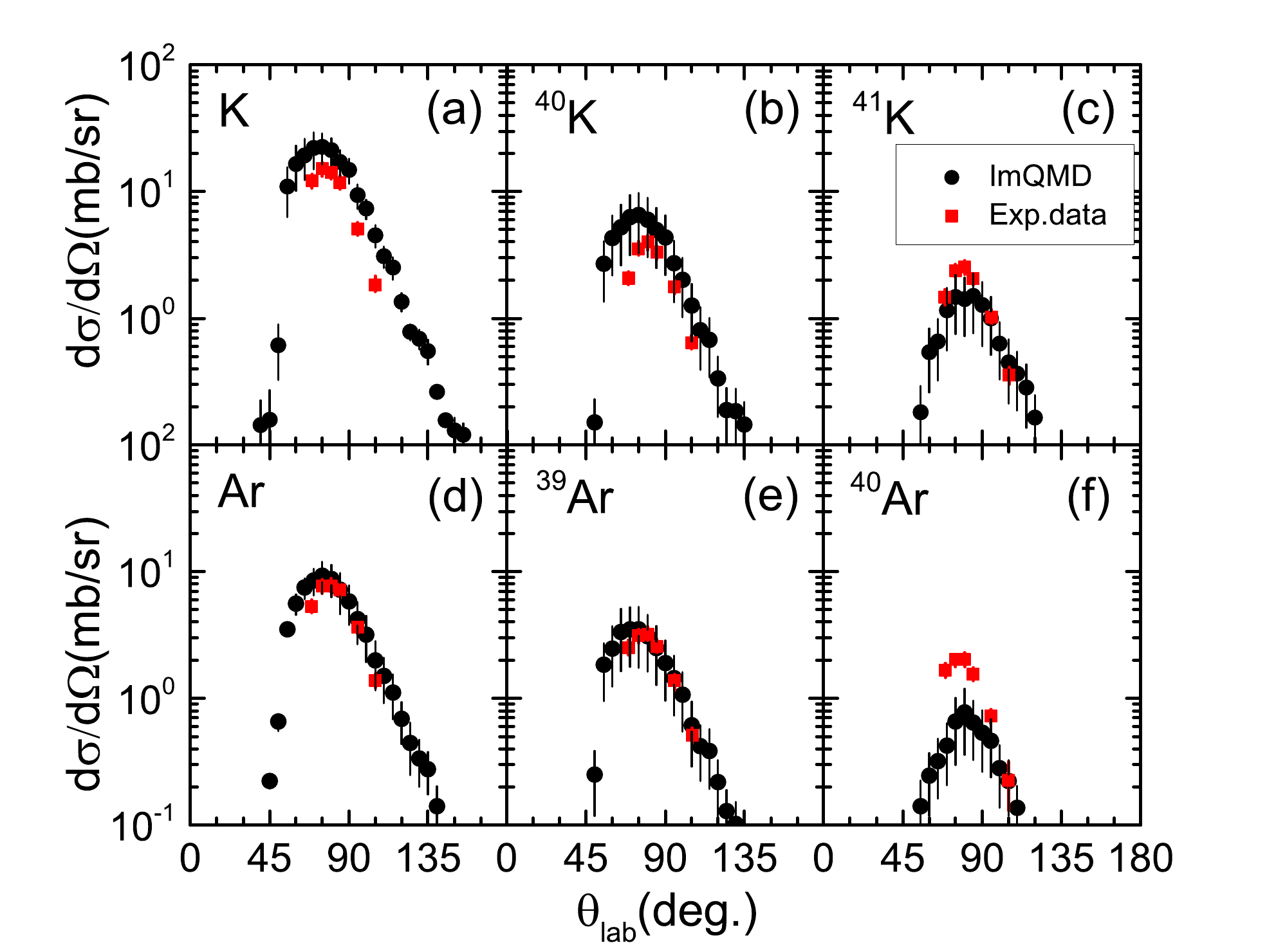}
\figcaption{\label{fig1} (Color online) Angular distributions of the final PLFs with different transfer channels in $^{40}$Ca+$^{124}$Sn at $E_{\textrm{c.m.}}=128.5$ MeV. The experimental data are taken from Ref. \cite{CWKB1}.}

\end{center}

Figure 2 shows the production cross sections of the final PLFs in $^{40}$Ca+$^{124}$Sn at $E_{\textrm{c.m.}}=128.5$ MeV. The squares and folding lines denote the calculations of the ImQMD model and CWKB theory \cite{CWKB1} following evaporation, respectively. The experimental data are taken from Ref. \cite{CWKB1}. It should be pointed out that the measured isotopic cross sections have been obtained by integrating the angular distributions via a quasi-Gaussian fit. From Fig. 2, the ImQMD calculations are in a reasonable agreement with the corresponding experimental data. The discrepancies between the calculated and experimental data are within one order of magnitude in general. One sees that the CWKB calculations reproduce the experimental data very well at the neutron-rich side of the distributions for the proton stripping channels from 0p ($Z=20$) to -4p ($Z=16$). However, it grossly underestimates the cross sections by several orders of magnitude at the neutron-deficient side. We don't show the calculations about the target-like-fragments (TLFs) because of the absence of experimental data. These results indicate that the ImQMD model is applicable for the study of MNT reactions of the intermediate-mass systems.

\begin{center}
\includegraphics[width=16cm]{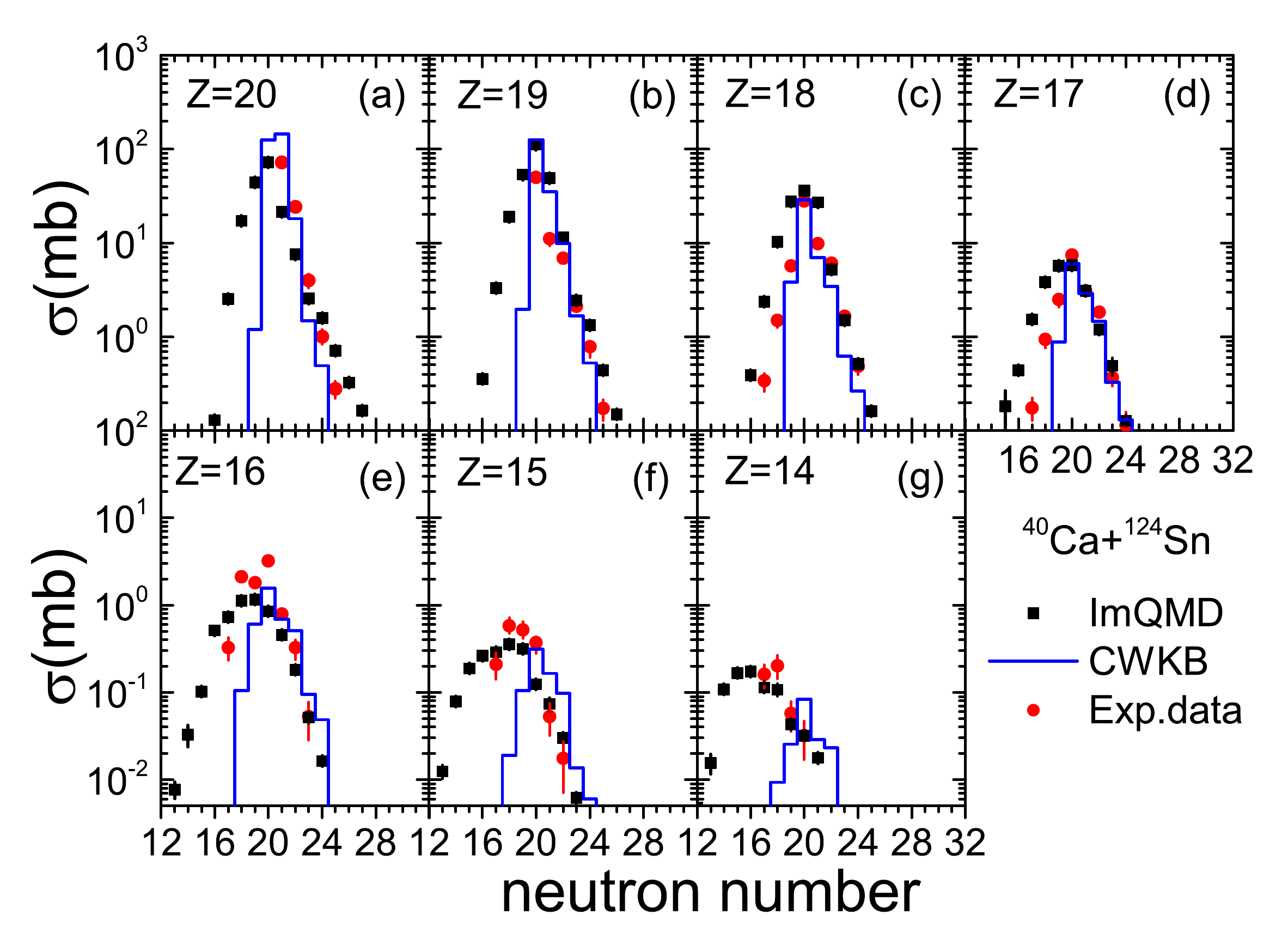}
\figcaption{\label{fig2} (Color online) Production cross sections of the final PLFs in $^{40}$Ca+$^{124}$Sn at $E_{\textrm{c.m.}}=128.5$ MeV. The black squares and folding lines denote the calculations of the ImQMD model and CWKB theory \cite{CWKB1} following evaporation, respectively. The experimental data are taken from Ref. \cite{CWKB1}.}

\end{center}

In order to produce the neutron-deficient nuclei near $N, Z$ = 50, choosing a favorable projectile-target combination is very important. Considering that the production cross sections at the maximum of the isotopic distributions decrease rapidly with increasing proton transfer channel, we choose the neutron-deficient nuclei, $^{106}$Cd and $^{112}$Sn, as targets. Figure 3 shows the calculated production cross sections of final TLFs with charge number from $Z=50$ to 54 in reactions of $^{48}$Ca+$^{112}$Sn, $^{40}$Ca+$^{112}$Sn, $^{58}$Ni+$^{112}$Sn and $^{106}$Cd+$^{112}$Sn. One sees that the production cross sections of the exotic neutron-deficient nuclei are the smallest in $^{48}$Ca+$^{112}$Sn system, hence the system is not suitable to produce such nuclei. For $^{40}$Ca+$^{112}$Sn, $^{58}$Ni+$^{112}$Sn, and $^{106}$Cd+$^{112}$Sn reactions, one can see that the discrepancies of the cross sections in the neutron-rich side are very significant. This is because the cross section in the neutron-rich side is very sensitive to $N/Z$ value of the projectile. The $N/Z$ values for $^{40}$Ca, $^{58}$Ni, and $^{106}$Cd are 1.00, 1.07, and 1.21, respectively. Therefore, the production cross sections of neutron-rich isotopes in $^{106}$Cd+$^{112}$Sn are larger than those in other two reactions. While in the neutron-deficient side, the isotopic distributions are similar for the three systems. This is because the primary distributions of three systems are almost the same in the neutron-deficient side.


\begin{center}
\includegraphics[width=16cm]{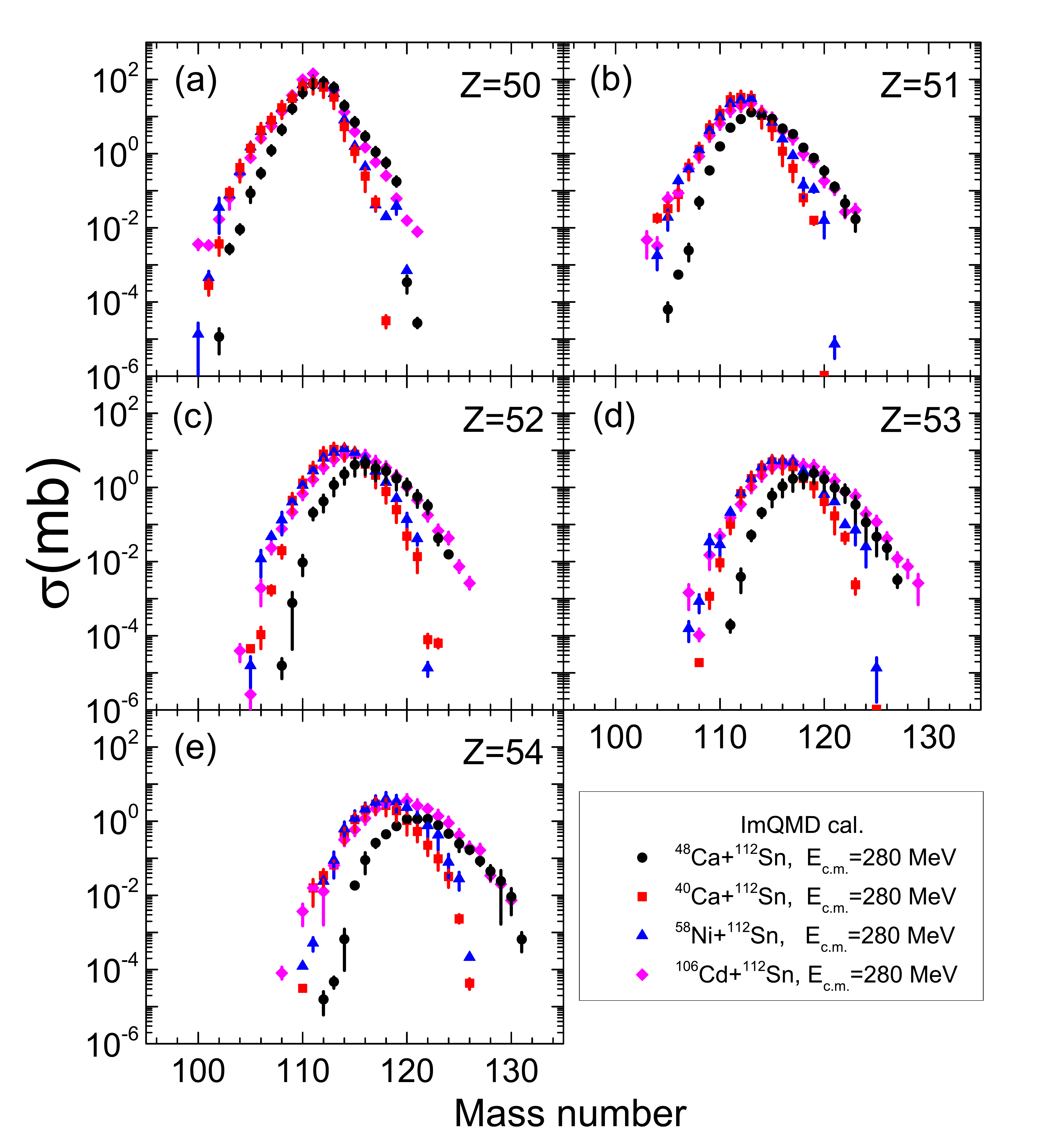}
\figcaption{\label{fig3} (Color online) Calculated production cross sections of final TLFs with charge number from $Z=50$ to 54 in different reactions at $E_{\textrm{c.m.}}=280 $ MeV.}

\end{center}

In MNT reactions, the energy dissipation process is a complex issue. The excitation energy of the reaction products is very important in their deexcitation processes. In the ImQMD model, the excitation energy of an excited fragment is calculated as $E^*=E_{\textrm{tot}}-E_\textrm{b}$. Here, $E_{\textrm{tot}}$ and $E_\textrm{b}$ denote the total energy and binding energy in ground state, respectively. The total energy of a fragment is the sum of all nucleon's kinetic energy in the body frame and potential energy. Figure 4(a) shows the average excitation energy of the products in binary events as a function of the impact parameters in $^{106}$Cd+$^{112}$Sn at $E_{\textrm{c.m.}}=500$ MeV. In the region of $b\geq 6$ fm, the total excitation energy of system increases rapidly with decreasing impact parameters. And it reaches a saturation value (about 145 MeV) at $b<5$ fm. The decay properties of the nuclei near the $^{100}$Sn are markedly different with neutron-rich nuclei. The emissions of proton and $\alpha$ particles were observed in the experiment in the decay processes of some specific nuclei even at their ground state. For deexcitation processes of these nuclei at excited states, the decay channels would be more complex. Figure 4(b) shows the yields of $n$, $p$, $d$, $t$, $^3$He, and $\alpha$ particles as a function of impact parameters in $^{106}$Cd+$^{112}$Sn system at $E_{\textrm{c.m.}}=500$ MeV. One can see that the charged particles emission plays an important role at small impact parameters in the de-excitation processes of the system. The number of emitted protons are much greater than that of other emitted charged particles. In the GEMINI simulation, we find that the protons emission is the main decay channel for the exotic neutron-deficient nuclei. In addition, the yields of the $\alpha$ and $d$ particles are considerable at small impact parameters. It results in a decrease of the yields of neutrons with the decrease of impact parameter at $b<6$ fm. Figure 4(c) shows the cross sections for formation of iodine isotopes (Z=53) in collisions $^{106}$Cd+$^{112}$Sn at $E_{\textrm{c.m.}}=500$ MeV. The open squares denote the unknown proton-rich nuclei. One can see that the final yields shift to the neutron-deficient side after the de-excitation process. More neutrons are evaporated on the neutron-rich side than on the neutron-deficient side after the deexcitation process. This is because neutron emission is the dominant decay channel for neutron-rich nuclei.
\begin{center}
\includegraphics[width=16cm]{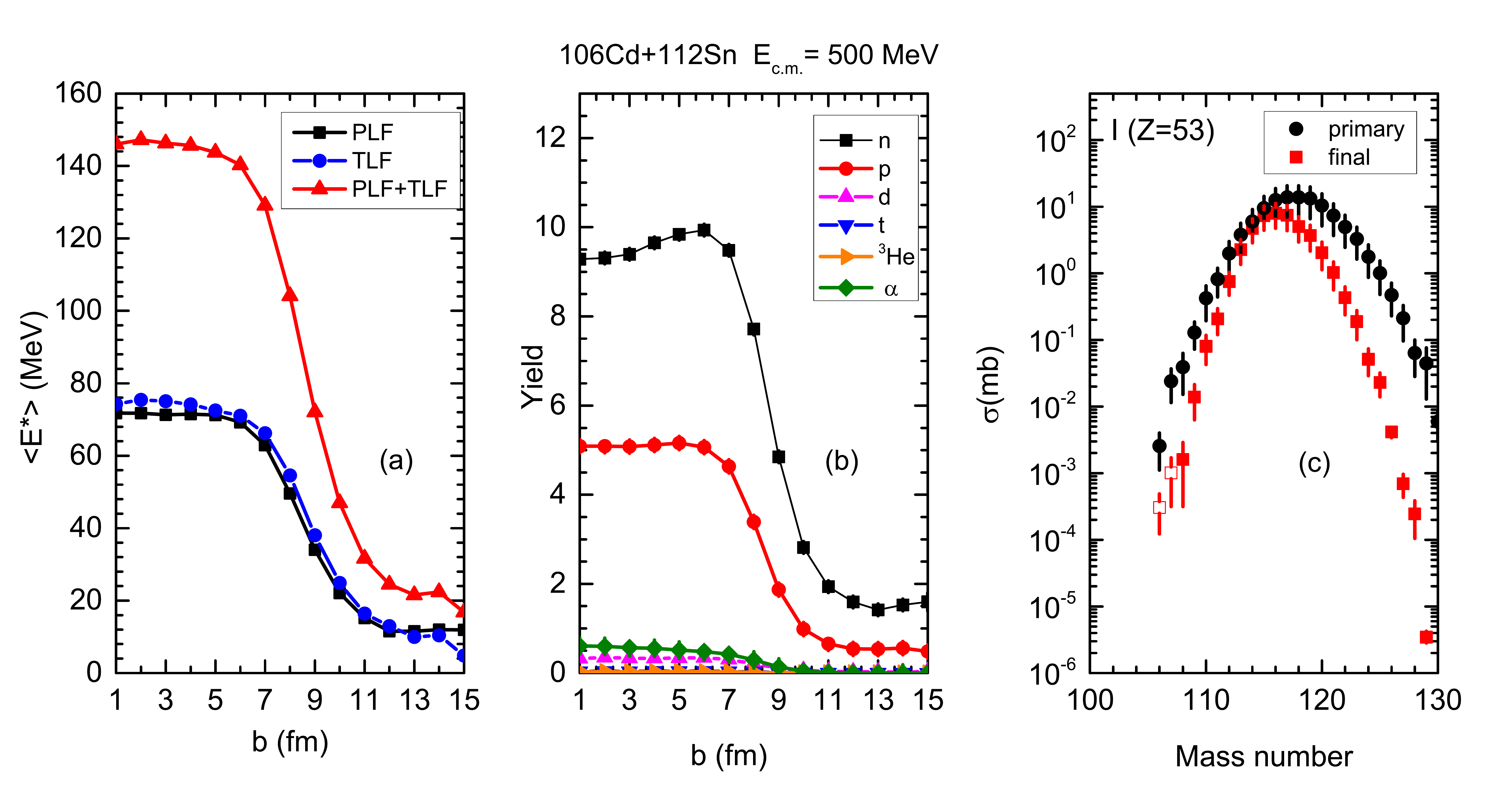}
\figcaption{\label{fig4} (Color online) (a) Average excitation energy of the products in binary events as a function of impact parameters in $^{106}$Cd+$^{112}$Sn at $E_{\textrm{c.m.}}=500$ MeV. (b) The yields of $n$, $p$, $d$, $t$, $^3$He, and $\alpha$ particles as a function of impact parameters in $^{106}$Cd+$^{112}$Sn system at $E_{\textrm{c.m.}}=500$ MeV. (c) Cross sections for formation of iodine isotopes (Z=53) in collision $^{106}$Cd+$^{112}$Sn at $E_{\textrm{c.m.}}=500$ MeV. The solid circles and squares denote distribution of primary and final fragments, respectively. The open squares denote the unknown neutron-deficient isotopes.}

\end{center}

Figure 5 shows the calculated isotopic distributions of final TLFs with charge number from $Z=50$ to 54 by the ImQMD model in the reactions of $^{106}$Cd+$^{112}$Sn at $E_{\textrm{c.m.}}=300$, 500, and 780 MeV. One sees that the isotopic production cross sections in the neutron-rich side become lower with increasing incident energies. The neutron evaporation is the main decay channel for the primary neutron-rich products. In the case of larger incident energy, it causes a larger shift of final distributions to the neutron-deficient side. While in the neutron-deficient side, the isotopic distributions are almost the same for these three incident energies. In general, larger incident energy improves the transfer probability of nucleons, which leads to larger production cross section for the primary neutron-deficient nuclei. However, the survival probability of these nuclei is lower due to higher excitation energies. Note that if the incident energy continues to increase, the production cross sections of the exotic neutron-deficient nuclei should be reduced, which is because the reactions are dominated by fragmentation mechanisms.
\begin{center}
\includegraphics[width=16cm]{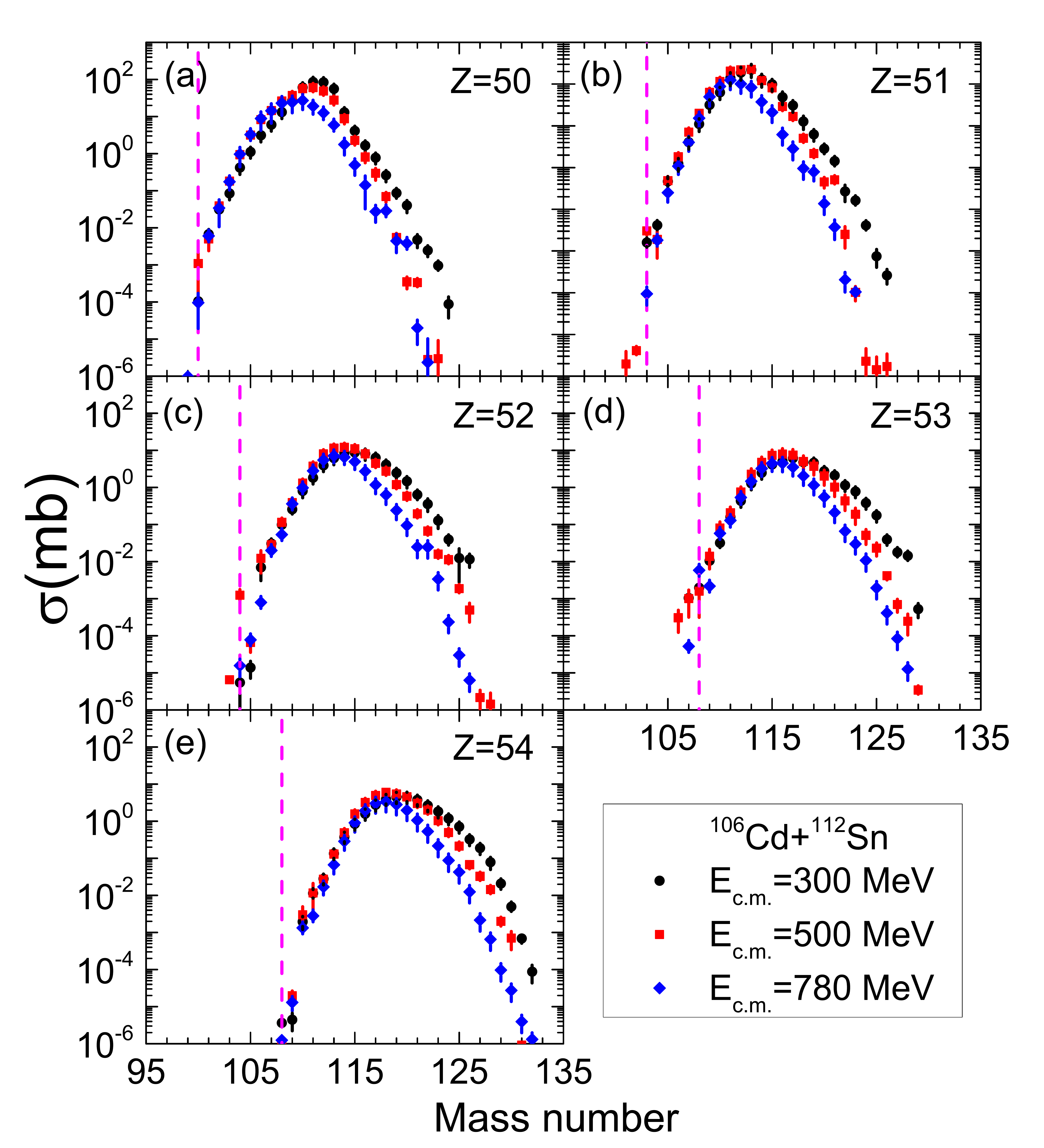}
\figcaption{\label{fig5} (Color online) Calculated isotopic distributions of final fragments by the ImQMD model in reactions of $^{106}$Cd+$^{112}$Sn at $E_{\textrm{c.m.}}=300$, 500, and 780 MeV. The dashed lines indicate the boundaries of known isotopes.}

\end{center}

Table 2 shows the comparison of the calculated production cross sections for exotic neutron-deficient nuclei from the MNT reactions with the measured values from the fragmentation and fusion-evaporation reactions. The measured cross sections from the fragmentation method are obtained in the reaction of 345 MeV/A $^{124}$Xe+Be \cite{Sn2,Sn6}. For the fusion-evaporation method, the cross sections of $^{100}$In and $^{101}$Sn are measured in $^{58}$Ni+$^{58}$Ni \cite{fus2} at $E_{\textrm{lab}}=348$ MeV; $^{100}$Sn is measured in $^{50}$Cr+$^{58}$Ni \cite{fus3} at $E_{\textrm{lab}}=255$ MeV; $^{108}$I, $^{109}$Xe, and $^{110}$Xe are measured in $^{58}$Ni+$^{54}$Fe \cite{fu1} at $E_{\textrm{lab}}=255$, 200, and 215 MeV, respectively. The cross sections of these isotopes from the MNT method are calculated with $^{106}$Cd+$^{112}$Sn at $E_{\textrm{c.m.}}=500$ MeV by using the ImQMD model. For the fusion-evaporation reactions, the production cross sections of the residual nucleus are one or two order of magnitude lower than those from the multinucleon transfer reactions. For projectile fragmentation, the production cross sections of these nuclei are much lower. For example, the cross sections of $^{100}$Sn by projectile fragmentation is only at the level of 10$^{-10}$ millibarn. Therefore, the MNT reactions have advantages in comparison to fusion-evaporation and projectile fragmentation reactions. Figure 6 shows the neutron-deficient nuclei region around $^{100}$Sn on the nuclear map. The filled and open squares denote the known and the predicted nuclei, respectively. Yellow, red and olive indicate the $\alpha$ decay, $\beta^{+}$ decay and proton decay, respectively. The production cross sections with ImQMD model in reaction of $^{106}$Cd+$^{112}$Sn at $E_{\textrm{c.m.}}=500$ MeV are signed in the graph. One can see that several new neutron-deficient nuclei are produced in the $^{106}$Cd+$^{112}$Sn reaction. The corresponding production cross sections for the new neutron-deficient nuclei, $^{101,102}$Sb, $^{103}$Te, and $^{106,107}$I, are 2.0 nb, 4.1 nb, 6.5 nb, 0.4 $\mu$b and 1.0 $\mu$b, respectively.
\begin{center}
\tabcaption{ \label{tab1} The comparison of the calculated production cross sections for exotic neutron-deficient nuclei from the MNT reactions with measured values from the fragmentation and fusion-evaporation reactions.}
\footnotesize
\begin{tabular*}{160mm}{c@{\extracolsep{\fill}}cccc}
\toprule Isotope & $\sigma^{\textrm{expt}}_{\textrm{frag}}$ (mb) & $\sigma^{\textrm{expt}}_{\textrm{fus}}$ (mb) & $\sigma^{\textrm{theo}}_{\textrm{MNT}}$ (mb) \\
			\hline
$^{97}$In & $1.3\times10^{-10}$\cite{Sn6}&--     &  $7.2\times10^{-6}$\\
$^{98}$In & $1.4\times10^{-8}~$\cite{Sn6}&--    &$3.5\times10^{-3}$  \\
$^{99}$In & $2.2\times10^{-7}~$\cite{Sn2}&--    & $2.0\times10^{-2}$  \\
$^{100}$In & $8.6\times10^{-6}~$\cite{Sn2}&$1.7\times10^{-3}$\cite{fus2}  & $3.8\times10^{-2}$ \\
$^{100}$Sn & $7.4\times10^{-10}$\cite{Sn2}&$4.0\times10^{-5}$\cite{fus3}  &$1.1\times10^{-3}$   \\
$^{101}$Sn & $4.0\times10^{-8}~$\cite{Sn2}&$1.3\times10^{-5}$\cite{fus2} &$5.0\times10^{-3}$   \\
$^{102}$Sn & $2.2\times10^{-6}~$\cite{Sn2}&-- &$3.9\times10^{-2}$   \\
$^{103}$Sn & $7.7\times10^{-5}~$\cite{Sn2}&-- &$1.8\times10^{-1}$   \\

$^{104}$Sb & $3.5\times10^{-7}~$\cite{Sn2}&--  & $1.9\times10^{-3}$  \\
$^{105}$Te & $1.2\times10^{-9}~$\cite{Sn2}&--  &$6.6\times10^{-5}$    \\
$^{106}$Te & $1.3\times10^{-7}~$\cite{Sn2}&-- & $1.2\times10^{-2}$  \\
$^{108}$I  & --                 &   $8.6\times10^{-4}$\cite{fu1}  &$1.6\times10^{-3}$  &    \\
$^{109}$Xe & --                 &   $1.0\times10^{-5}$\cite{fu1} & $2.0\times10^{-5}$   \\
$^{110}$Xe & --                 &   $1.0\times10^{-3}$\cite{fu1} & $3.0\times10^{-3}$   \\
\bottomrule
\end{tabular*}
\end{center}

\begin{center}
	\includegraphics[width=16cm]{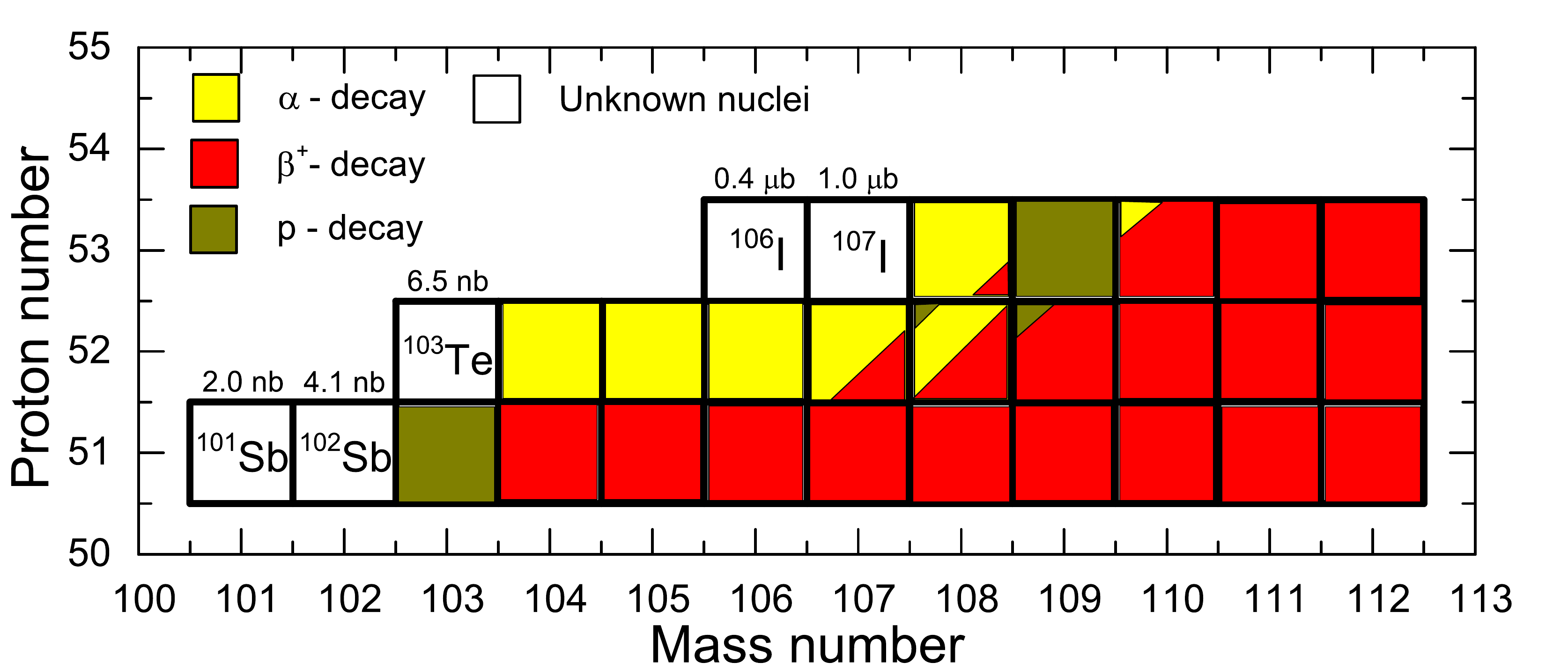}
\figcaption{\label{fig6} Neutron-deficient nuclei region around $^{100}$Sn on the nuclear map. The filled and open squares denote the known and predicted nuclei, respectively. Yellow, red and olive indicate the $\alpha$ decay, $\beta^{+}$ decay and proton decay, respectively. The production cross sections in reaction of $^{106}$Cd+$^{112}$Sn at $E_{\textrm{c.m.}}=500$ MeV are signed in the graph.}	\label{fig6}
\end{center}

\section{Conclusions}
The production cross sections and the angular distributions of PLFs in the reaction of $^{40}$Ca+$^{124}$Sn at $E_{\textrm{c.m.}}=128.5$ MeV are calculated by the ImQMD model. The results show that the ImQMD model is suitable to describe the MNT reactions of the intermediate-mass systems. In order to produce the exotic neutron-deficient nuclei around $N, Z$ = 50, the multinucleon transfer reactions of $^{48}$Ca+$^{112}$Sn, $^{40}$Ca+$^{112}$Sn, $^{58}$Ni+$^{112}$Sn and $^{106}$Cd+$^{112}$Sn are studied. We find that combinations of neutron-deficient projectile and target are advantageous for the production of the exotic neutron-deficient nuclei. The distribution of the final production cross section of exotic neutron-deficient nuclei are similar in the reactions of $^{106}$Cd+$^{112}$Sn at $E_{\textrm{c.m.}}=300$, 500, and 780 MeV. The deexcitation processes of $^{106}$Cd+$^{112}$Sn system at $E_{\textrm{c.m.}}=500$ MeV are analysed. It is found that the charged particles emission plays an important role for the highly excited system.
Protons emission is the main decay channel in the deexcitation processes of exotic neutron-deficient nuclei. Compared to the fragmentation and fusion-evaporation reactions, the MNT reaction is very advantageous for the production of the neutron-deficient nuclei around $N, Z$ = 50. The cross sections of unknown neutron-deficient isotopes $^{101,102}$Sb, $^{103}$Te, and $^{106,107}$I are 2.0 nb, 4.1 nb, 6.5 nb, 0.4 $\mu$b and 1.0 $\mu$b in $^{106}$Cd+$^{112}$Sn reaction, respectively.


\vspace{10mm}

\vspace{-1mm}
\centerline{\rule{180mm}{0.1pt}}
\vspace{2mm}




\clearpage

\end{document}